# Experimental and theoretical justifications for observed discriminations between enantiomers of prochiral alcohols by chirally blind EI-MS

Mohamad Z. Kassaee[a,b,c*], Ahmad Mani-Varnosfaderani[b*], Nastaran Abedini[b], Esmaeil Eidi[b], Aysan Mojtaheadi[b], Mohamad H. Kassaee[c], Peter T. Cummings[a*]

[a]Chemical and Biomolecular Engineering, Vanderbilt University, Nashville, TN 37240.

[b]Chemistry Department, Tarbiat Modares University, 14115-175, Tehran, Iran.

[c]Pure and Applied Research LLC, P.O. Box 111591, Nashville, Tennessee 37222

*Correspondence

Email: kassaeem@modares.ac.ir

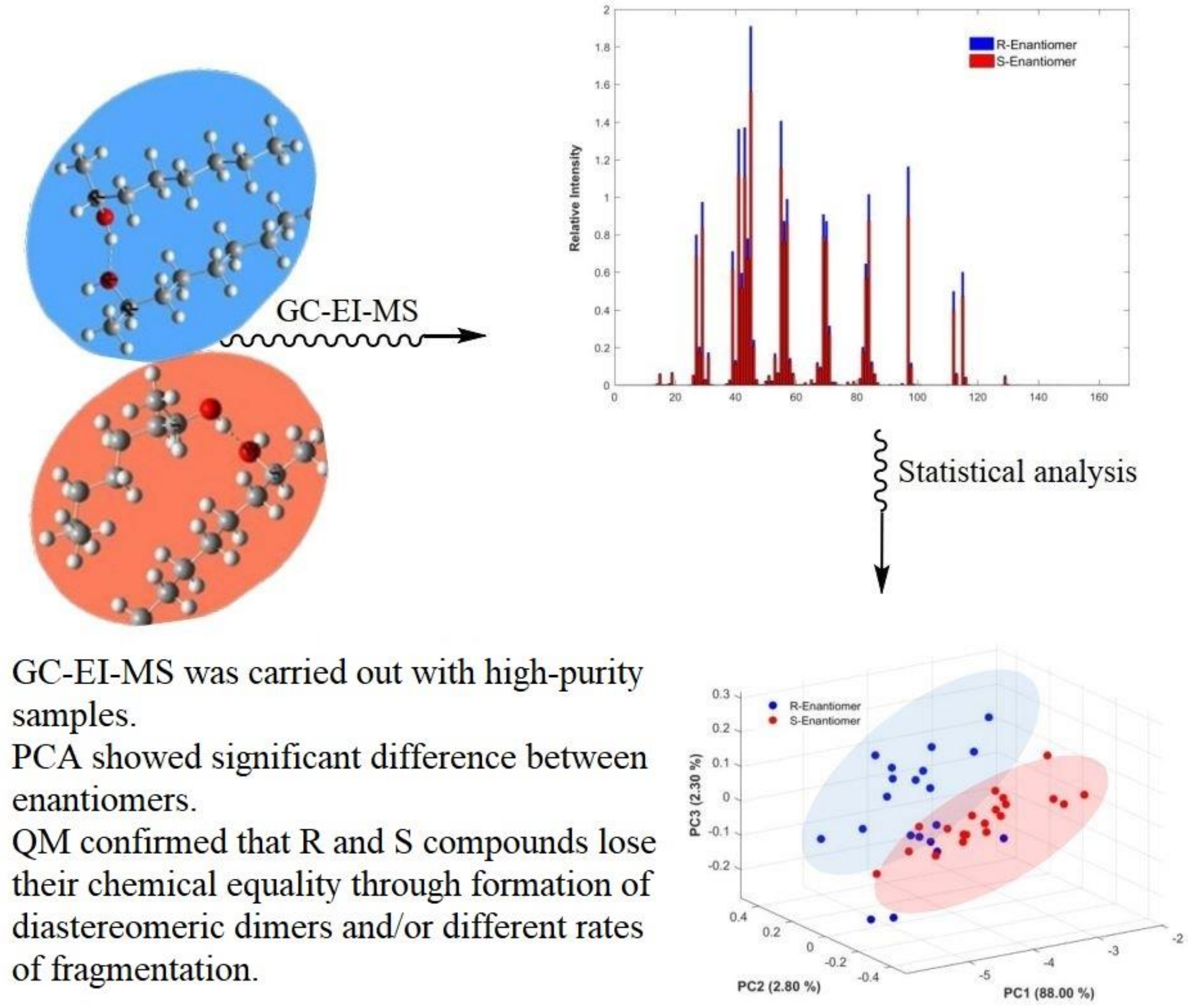

**Abstract**

To all appearances, electron impact mass spectrometer (EI-MS) is considered a ‘‘chirally blind’’ instrument. However, numerous non-identical **R** (right handed) and **S** (left handed) configurations of prochiral alcohols' mass spectra have appeared in the literature with almost no justification. Such observations are often attributed to impurities, experimental circumstances, inaccurate measurements, *etc*. In an experimental attempt to explain this phenomenon, here we have avoided the above mentioned pitfalls by conducting control experiments using different pure enantiomers under the same circumstances. Hence, we report the mass spectra of ***R***- and ***S***-enantiomers of 2-octanol ($\mathbf{1}_{\boldsymbol{R}}$, $\mathbf{1}_{\boldsymbol{S}}$) and 1-octyn-3-ol ($\mathbf{2}_{\boldsymbol{R}}$, $\mathbf{2}_{\boldsymbol{S}}$) collected by running 20 independent experiments for each ***R***- and ***S***-enantiomer. Statistical analyses confirmed that the peak intensities were significant to an acceptable level of confidence. The $\mathbf{1}_{\boldsymbol{R}}$ and $\mathbf{1}_{\boldsymbol{S}}$ enantiomers were separated reasonably in the PC space, implying that the chirally blind EI-MS is able to discriminate between **R** and **S** prochiral alcohols. Theoretically, self-complexation through H-bonding for **S** (or **R**) appears to give a new chiral center at the H-bonded oxygen atom, producing a new dimeric pair of diastereomers **SRS**$^{\cdot+}$ and **SSS**$^{\cdot+}$ (or **RRR**$^{\cdot+}$ and **RSR**$^{\cdot+}$) after ionization. The calculation results of Gibbs free energies ($\Delta G < 0$) and the equilibrium distribution ($K_{eq} > 1$) have shown that these hydrogen bonds are formed. Interestingly, the latter four ionized diastereomers appear with different structural and thermodynamic parameters at the M06-2X/6-311++g (d,p) level of theory.

**Keywords:** EI-MS spectra, Prochiral, Enantiomer, Mass spectra.

## Introduction

Chiral discrimination is vital in various fields.[1-4] Most medically important compounds are chiral with only one of their enantiomers exhibiting biological activity and the other being often inactive with harmful effects in a chiral environment, such as in biological systems.[5] The current trend in medicinal advancement and the market is increasingly moving towards developing pure enantiomeric drugs for their therapeutic values, such as reducing the required dose, increasing the potency, and improving their safety profile.[5] The food and drug administration (FDA) guidelines require knowledge of chirality for compounds with chiral centers for *in vitro* and *in vivo* studies during the drug development.[6] Traditionally, this information is not solicited from analytical tools like an electron impact mass spectrometer (EI-MS) which lacks an asymmetric environment.[7-11] An asymmetric environment is required to distinguish between non-superimposable enantiomers including nearly all amino acids, sugars, proteins, nucleic acids, lipids, carbohydrates, agrochemicals, foods, pharmaceuticals, and many industrial compounds. Detection methods are mainly based on diastereomer formation *via* chiral substrates, chiral selectors, chiral metal complexes, *etc*.[12-17] Discrimination between enantiomers is customarily achieved by polarimetry, optical rotator dispersion (ORD), circular dichroism (CD), *etc*.[18-21] However, many biochemical interactions in the body are based on non-bonding interactions of chiral and prochiral structures. These interactions can alter the properties of chiral compounds by creating new chiral centers. Hence, formation of temporary diastereomers may explain these changes.[22-27] In this work, we show a unique potential of a routine chirally blind gas chromatograph EI-MS instrument in distinguishing between alcohols containing prochiral centers.

## Experimental and Computational Methods

### 1. Experimental investigation

The GC–MS analyses were performed with the Agilent 5973 inert Gas Chromatograph/Mass Spectrometer (Ionization mode justified on EI, Ionization energy 70 eV) with a transfer line temperature hold at ambient temperatures; helium was used as the carrier gas (constant flow, 1 mL/min). The ion source temperature was fixed at 50 °C to maintain possible intermolecular interactions in the studied structures. Chromatographic separations were performed with an Agilent J&W DB-5ms column, which is a nonpolar phenyl arylene polymer, virtually equivalent to (5%-phenyl)-methylpolysiloxane (30 m, 0.25 mm film thickness 1.40 μm). The oven temperature was set at 50 °C. Total ion chromatograms (TICs) were recorded (full scan range from m/z 15 to m/z 160, at 1.05 scan $s^{-1}$).

The PCA approach was used for dimension reduction and visualization of the collected data in this work. We used MATLAB software (version 2022a) for implementation of the PCA approach and the singular value decomposition (SVD) algorithm was used for this purpose. The Fisher's least significant difference (LSD) at the 0.05% of statistical probability (*e.g.* $P \leq 0.05$- means with 95% accuracy) was used for determining the level of meaningful differences between the averages of mass spectra. The LSD calculates the smallest significant difference between means from different individual groups. Any difference larger than the LSD is considered a significant result. The LSD calculations were conducted using the multcompare function in MATLAB software (version 2022a). The two-sided T-test and F-test analyses (called t value in student's t test and F value in ANOVA (Analysis of Variance) test) were conducted using the facilities provided in both MATLAB and Microsoft Excel software tools.

**Theoretical investigation**

Systematic conformer search for complexes were carried out using Spartan'14 and the molecular mechanics (MMFF94).[28-31] All quantum chemical calculations were performed using Gaussian 09 suite of programs.[32] Geometry optimizations and frequency calculations of studied molecules were achieved using M06-2X hybrid density functional theory calculations with Pople type 6-311++g (d, p) basis set, which is the most efficient method to describe the *van der* Waals interactions at a relatively small computational cost.[33,34] Interaction energy and Gibbs free energy were calculated following equation (1)

$$\Delta\Delta G = \Delta G_{AB} - \Delta G_{A} - \Delta G_{B} \qquad (1)$$

To overcome basis set superposition errors (BSSE), we used counterpoise (CP) method to ensure the simultaneous optimization of geometries of dimeric complexes.[35,36] This method is prevalent and validated in many other papers on alcohols.[37-44]

We can calculate the equilibrium distribution of each species in the mixture on the basis of its equilibrium constant (2)

$$K_{eq} = \frac{N_1}{N_2} = \exp\left(-\frac{\Delta G}{RT}\right) \qquad (2)$$

Where ΔG is the difference in Gibbs free of respective species at 298.15 K.[45]

## Results and Discussion

### 1. Experimental

The EI-MS appeared in 1900's, with an undesirable classical claim that all "enantiomers give an identical mass spectrum under EI-MS achiral operating conditions"[11]. To all appearances, EI-MS is still considered a ''chirally blind'' instrument. Yet, numerous non-identical *R* and *S* mass spectra of enantiomeric prochiral pairs are found in the literature. As samples, we show ten clearly different spectra of *R* and *S* pairs, obtained from the library of mass spectra NIST web book (Supplementary 1). This phenomenon may be attributed to impurities, experimental circumstances, inaccurate measurements, *etc*. Here, we have avoided such pitfalls by conducting control experiments using pure enantiomers under the same circumstances. We provide statistical analysis to confirm that the peak intensities for the scrutinized enantiomers are significant to an acceptable level of confidence.[46] For example, we report the mass spectra collected by running 20 independent experiments for each enantiomer ($\mathbf{1}_{\boldsymbol{R}}$ and $\mathbf{1}_{\boldsymbol{S}}$), which were internally normalized concerning the signal recorded at m/z = 54. The latter was selected for the normalization procedure since it was not saturated in all experiments. Moreover, several different internal normalizations were performed concerning other m/z values, but the same results were obtained (data not shown). The internal normalization procedure eliminates the effects of different random variables, which may affect the concentration of analytes in the mass analyzer. In order to explore the similarities of signals for different samples and enantiomers, principle component analysis (PCA) has been performed on collected data. The PCA approach reduces the dimension of this multivariate data and provides a systematic way for visualization purposes. The first principle components (PCs) describe most variances of the signals, while the last PCs are attributed to the distribution of noise in the data. The plot of the first PCs against each other represents the contribution of meaningful signals for describing the data. The first three PCs in this work cover 95.5% of the data variances. The plot of the PC1 against PC2 and PC3 is shown in a 3D-plot (Fig. 1a, top). Evidently, most of the $\mathbf{1}_{\boldsymbol{R}}$ and $\mathbf{1}_{\boldsymbol{S}}$ enantiomers are separated reasonably in the PC space. This score plot implies that the mass spectra for $\mathbf{1}_{\boldsymbol{R}}$ and $\mathbf{1}_{\boldsymbol{S}}$ enantiomers are significantly different. The average mass spectra for $\mathbf{1}_{\boldsymbol{R}}$ and $\mathbf{1}_{\boldsymbol{S}}$ enantiomers show the $\mathbf{1}_{\boldsymbol{R}}$ peaks are more intense than the $\mathbf{1}_{\boldsymbol{S}}$ signals (Fig. 1b, top). Mass spectra average differences appear neither the same nor homogeneous for discriminatory m/z channels of $\mathbf{1}_{\boldsymbol{R}}$ and $\mathbf{1}_{\boldsymbol{S}}$. In contrast, the magnitude of lots of m/z channels is the same in the two enantiomers (Fig. 1c, top). The student's t-test has been used to compare the average values of signals for different m/z channels of $\mathbf{1}_{\boldsymbol{R}}$ and $\mathbf{1}_{\boldsymbol{S}}$ enantiomers. The most discriminatory m/z channels, with the most different values of signals for $\mathbf{1}_{\boldsymbol{R}}$ and $\mathbf{1}_{\boldsymbol{S}}$ enantiomers, are listed (Supplementary 2). The calculated p-values reveal that the observed differences between mass peaks for $\mathbf{1}_{\boldsymbol{R}}$ and $\mathbf{1}_{\boldsymbol{S}}$ enantiomers are significant. Likewise, differences between mass peaks for $\mathbf{2}_{\boldsymbol{R}}$ and $\mathbf{2}_{\boldsymbol{S}}$ enantiomers appear significant, under similar treatments (Fig. 1a-c, bottom; Supplementary 3). Moreover, to confirm the accuracy of the data obtained in the conformational energy calculations and show significant differences, we performed T-student test analyses on the dimer data set of the studied molecules.

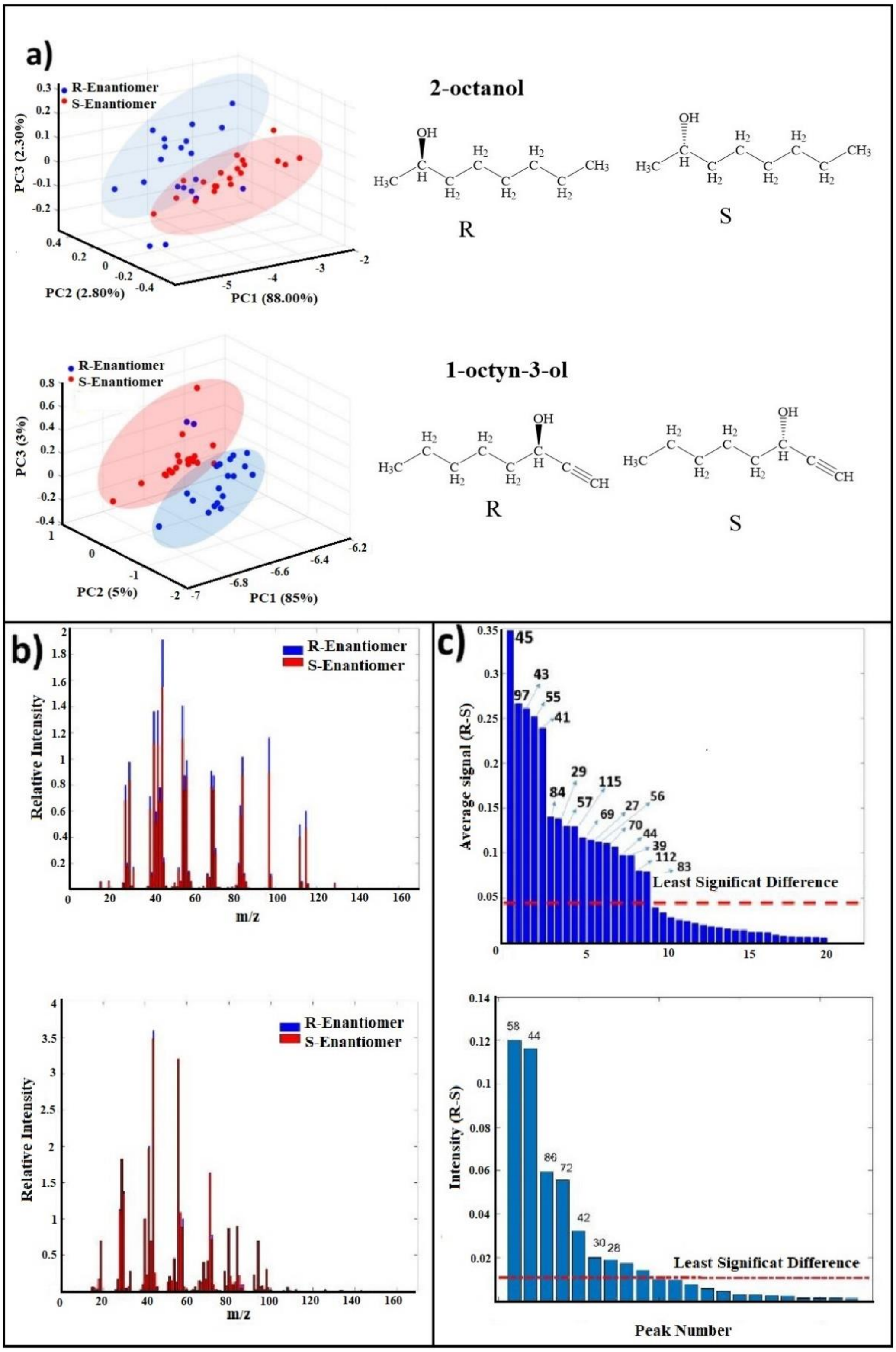

**Fig. 1**. Statistical analysis for enantiomers of 2-octanol (**1***R*, **1***S*; top), along with those of 1-octyn-3-ol (**2***R*, **2***S*; bottom) a) A 3D-plot of the PC1 against PC2 and PC3. b) The average mass spectra. c) Average differences between the mass spectra.

Interestingly, the GC-chromatogram of pure $\mathbf{1}_{\boldsymbol{R}}$ and $\mathbf{1}_{\boldsymbol{S}}$ exhibit several small peaks at higher retention times, which show the same m/z $\mathbf{1}_{\boldsymbol{R}}$ and $\mathbf{1}_{\boldsymbol{S}}$ (Fig. 2 b-c). Although the existence of hydrogen dimers in alcohols has not been fully proven, several reports have confirmed the existence of these interactions.[47-49] Such associated molecules may give higher retentions because of their polar intermolecular interactions and/or non-polar ones which exhibit higher interaction strength for larger molecules.[50-52] Moreover, it has been established that in simulations of the secondary alcohols, the formation of assemblies as ring clusters take precedence over chain structures, and the most probable cluster size consists of two monomers with tails longer than 5 carbon.[53] Due to the fact that self-assembly does not follow a specific pattern at the gas phase, there is a high probability of forming several dimers with different conformers. Therefore, we used conformer search to find the most likely dimer and repeated 20 times.

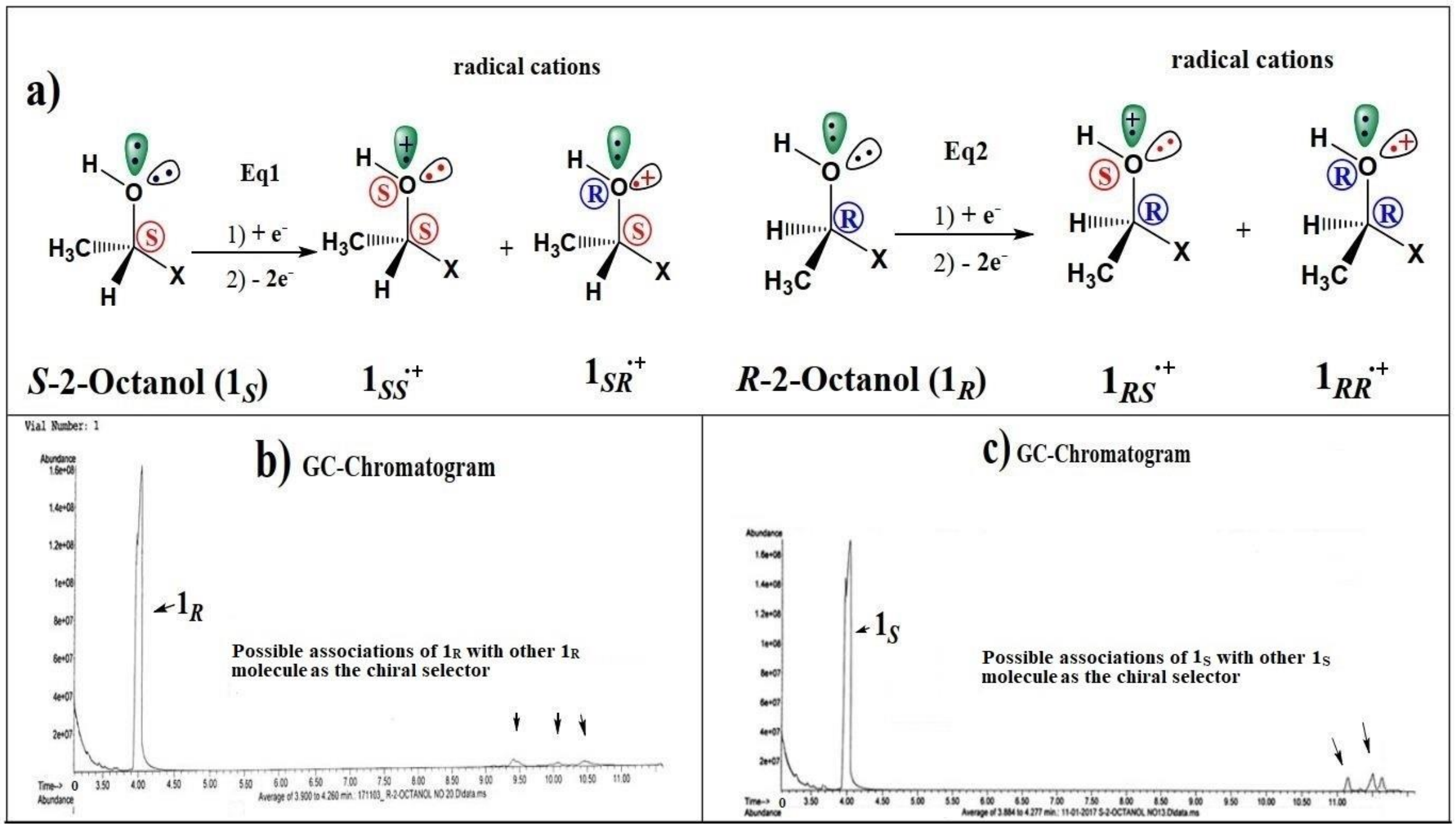

**Fig. 2.** a) Conversion of prochiral ***S***-2-octanol, $\mathbf{1}_{\boldsymbol{S}}$ (or its enantiomer ***R***-2-octanol, $\mathbf{1}_{\boldsymbol{R}}$) to the seemingly diastereomeric radical cations, $\mathbf{1}_{\boldsymbol{SS}}{}^{\cdot+}$ and $\mathbf{1}_{\boldsymbol{SR}}{}^{\cdot+}$(or $\mathbf{1}_{\boldsymbol{RS}}{}^{\cdot+}$ and $\mathbf{1}_{\boldsymbol{RR}}{}^{\cdot+}$) upon ionization

in EI-MS chamber, Eqs.1 (or Eqs 2) (Supplementary 5). b, c) Gas chromatogram of $\mathbf{1}_{\boldsymbol{R}}$ and $\mathbf{1}_{\boldsymbol{S}}$. Peaks at longer retention times show the same GC-MS spectras $\mathbf{1}_{\boldsymbol{R}}$ and $\mathbf{1}_{\boldsymbol{S}}$, respectively (Supplementary 6).

## 2. Theoretical

Interaction energies for H-bonded dimers were instructive (Fig. 3 and Table 1). It is the energy released when the H-bonded oxygen atom, producing a new dimeric pair of diastereomers $\mathbf{1}_{\boldsymbol{SRS}}^{\cdot+}$ and $\mathbf{1}_{\boldsymbol{SSS}}^{\cdot+}$ (or $\mathbf{1}_{\boldsymbol{RSR}}^{\cdot+}$ and $\mathbf{1}_{\boldsymbol{RRR}}^{\cdot+}$) after ionization. The results of our calculations with exothermic reactions with a negative ΔG (-10.64 to -11.24 kcal/mol) have explicitly shown that these hydrogen bonds formed. The discrimination between the enantiomers may be justified through their possible self-assemblies as chiral selectors, which could be made diastereomeric complexes *via* H-bonding. The reason is that it is, without doubt, the larger H-bond aggregates that the chirally blind EI-MS is able to discriminate between **R** and **S** prochiral alcohols.

Our calculations at M06-2X/6-311++g(d,p) level of theory confirmed our experimental findings. The dimers did not have the same structural properties, and the average energy of the dimers formed in $\mathbf{1}_{\boldsymbol{R}}$ was the same as $\mathbf{1}_{\boldsymbol{S}}$. Finally, we selected the most stable conformers and calculated their quantum properties. Interestingly, the non-polar and H-bonding association of $\mathbf{1}_{\boldsymbol{R}}$ (as a chiral selector) with an ionized form molecule of itself ($\mathbf{1}_{\boldsymbol{R}}^{\cdot+}$, after ionization) gives several diastereomers, whose most stable form is $\mathbf{1}_{\boldsymbol{RR}}^{\cdot+}$ with 10.64 kcal/mol less Gibbs free energy that the sum of two separated (most stable) conformers of $\mathbf{1}_{\boldsymbol{R}}$ and $\mathbf{1}_{\boldsymbol{R}}^{\cdot+}$ (Fig. 3c). A similar association of $\mathbf{1}_{\boldsymbol{S}}$ (as a chiral selector) with an ionized form molecule of itself ($\mathbf{1}_{\boldsymbol{S}}^{\cdot+}$, after ionization) gives several diastereomers, whose most stable form is $\mathbf{1}_{\boldsymbol{SS}}^{\cdot+}$ with 10.64 kcal/mol less Gibbs free energy than the sum of two separated (most stable) conformers of $\mathbf{1}_{\boldsymbol{S}}$ and $\mathbf{1}_{\boldsymbol{S}}^{\cdot+}$ (Fig. 3d). On the other hand, dimerization of $\mathbf{1}_{\boldsymbol{R}}$ and $\mathbf{1}_{\boldsymbol{S}}$ is endothermic giving $\mathbf{1}_{\boldsymbol{RR}}$ and $\mathbf{1}_{\boldsymbol{SS}}$ with 3.60 and 3.61 kcal/mol more Gibbs free energy than the sum of their monomeric constituents (Fig. 3a-b). To understand the strength of hydrogen bonds and possibly their ability to form a new chiral center, we carried out an AIM calculation on the most stable dimers at M06-2x/6-311++g(d,p) level of theory. Strong and moderate H-bonding interactions range from 15.0-40 and 4-15 kcal/mol, respectively.[54] The results of our calculations categorized the hydrogen bonds formed, especially in ionic dimers, as strong, showing 24.30 kcal/mol for $\mathbf{1}_{\mathbf{RR}}^{\cdot+}$ and $\mathbf{1}_{SS}^{\cdot+}$, plus 9.04 for $\mathbf{1}_{\mathbf{RR}}$ and $\mathbf{1}_{SS}$, respectively. Further experimental scrutiny of the latter is beyond the scope of this manuscript and will be accounted for elsewhere. The $\Delta G < 0$ and $K_{eq} > 1$ for $\mathbf{1}_{\mathbf{RR}}^{\cdot+}$, $\mathbf{1}_{SS}^{\cdot+}$, $\mathbf{2}_{\mathbf{RR}}^{\cdot+}$, and $\mathbf{2}_{SS}^{\cdot+}$ indicate that the H-bonding formations are favored (Table 1).

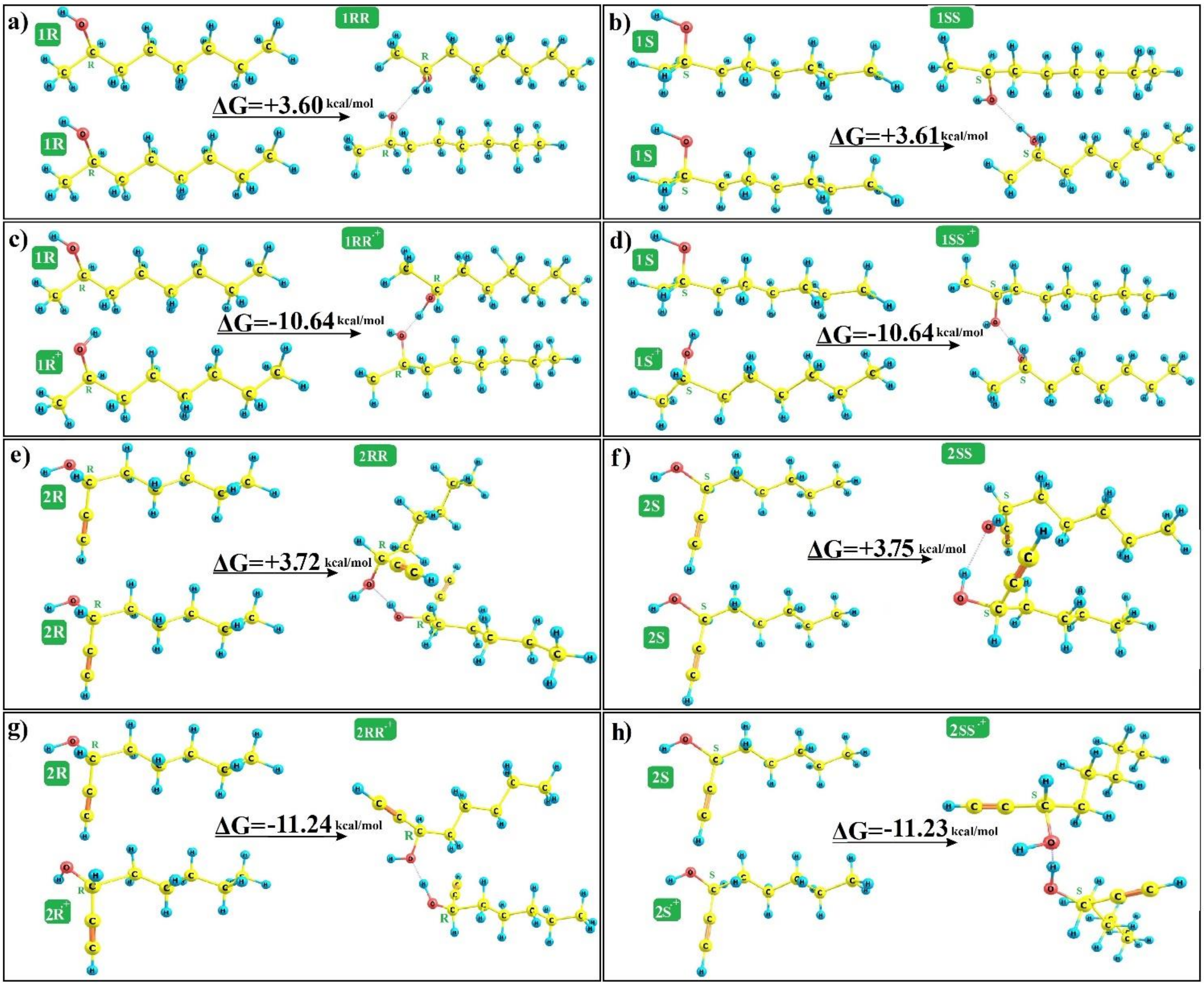

**Fig. 3.** Our calculations attribute spectral differences between Gibbs free energy (ΔG) of species (**1*R***, **1*S***, **2*R***, **2*S***) to the differences in their H-bonded dimers or the differing interactions with the resulting radical cations as chiral selectors. Selected energies and thermodynamic parameters from DFT optimized neutral species (**1*R***, **1*S***, **2*R***, **2*S***) and their diastereomeric dimers are shown in Supplementary 7**.**

**Table 1**. Super-molecular interaction (H-bonding dimerization) energies between unionized $^1E_A$ species *per se* (**1*R*, 1*S*, 2*R***, **2*S***), forming $^3E_{AA}$ dimers (**1*RR***, **2*RR***, **1*SS***, **2*SS***); along with interaction energies between $^1E_A$ species and $^2E_B$ radical cations (**1*R***$^{\cdot+}$, **1*S***$^{\cdot+}$, **2*R***$^{\cdot+}$, **2*S***$^{\cdot+}$), giving dimeric (**1*RR***$^{\cdot+}$, **1*SS***$^{\cdot+}$, **2*RR***$^{\cdot+}$, and **2*SS***$^{\cdot+}$), Gibbs free energy (ΔG), and equilibrium constant ($K_{eq}$) at M06-2X/6-311++g(d,p) level of theory.

| Monomers | Energy (a.u.) $^1E_A$ or $^2E_B$ | Dimers | Energy (a.u.) $^3E_{AA}$ or $E_{AB}$ | $\Delta E^a$ (kcal/mol) | $\Delta G^b$ (kcal/mol) | $K_{eq}$ |
|---|---|---|---|---|---|---|
| **1*R*** | -390.575587 | **1*RR*** | -781.16558 | -9.04 | 3.60 | $2.29*10^{-3}$ |
| **1*R***$^{\cdot+}$ | -390.230653 | **1*RR***$^{\cdot+}$ | -780.84497 | -24.30 | -10.64 | $6.33*10^{+7}$ |
| **1*S*** | -390.575587 | **1*SS*** | -781.16558 | -9.04 | 3.61 | $2.23*10^{-3}$ |
| **1*S***$^{\cdot+}$ | -390.230653 | **1*SS***$^{\cdot+}$ | -780.84497 | -24.30 | -10.64 | $6.32*10^{+7}$ |
| **2*R*** | -388.146536 | **2*RR*** | -776.302547 | -5.95 | 3.72 | $1.89*10^{-3}$ |
| **2*R***$^{\cdot+}$ | -387.797158 | **2*RR***$^{\cdot+}$ | -775.974193 | -19.14 | -11.24 | $1.73*10^{+7}$ |
| **2*S*** | -388.146536 | **2*SS*** | -776.307104 | -8.81 | 3.75 | $1.78*10^{-3}$ |
| **2*S***$^{\cdot+}$ | -387.797158 | **2*SS***$^{\cdot+}$ | -775.97875 | -22.00 | -11.23 | $1.72*10^{+7}$ |

$^a\Delta E = E_{AA}$ or $E_{AB} - (E_A+E_B)$

$^b\Delta G = G_{AA}$ or $G_{AB} - (G_A+G_B)$

In order to understand the bonding or non-bonding nature of interaction between the chiral centers and its effects on the stability of alcohol complexes, we carried out the quantum theory of atoms in molecules (QTAIM) calculations, using AIM2000[55] and Multiwfn[56] software packages (Fig. 4 and Table 2). Electron density values ($\rho_{BCP}$) of 0.110 and 0.009 for intermolecular interactions OH---O and CH----O between two monomers of $\mathbf{1_{RR}}^{\cdot+}$, as well as their corresponding Laplacian ($\nabla^2\rho_{BCP}$) of 0.100 and 0.030 indicate that the above mentioned interactions are H-bonding and *van der* Waals interactions, respectively.[57] In addition, to determine the contribution of such interactions to the stability of the complexes, we calculated interactions energy H-bonding ($\Delta E_{HB}$), which was equivalent to the average potential energy (V) at the bond critical point (BCP). The G and V are the kinetic and potential energy density at BCP, respectively.

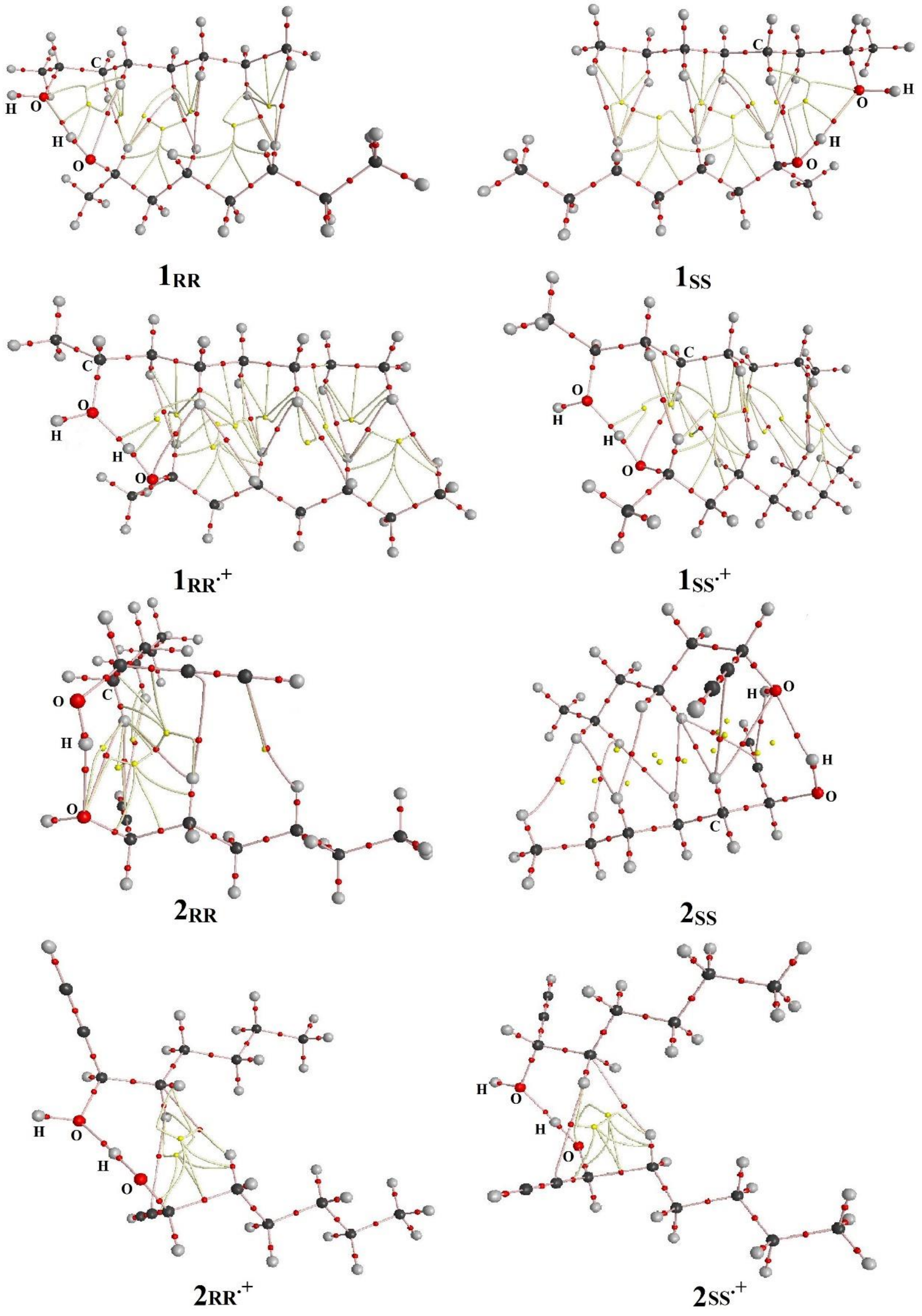

**Fig. 4.** Atoms in molecules (AIM) wave function analysis, showing bond critical points (BCP) with small red spheres; and the ring critical points (RCP) with very small yellow spheres.

**Table 2.** Electron density ($\rho_{BCP}$), Laplacian ($\nabla^2\rho_{BCP}$), potential energy ($V_{BCP}$), kinetic energy ($G_{BCP}$), total electron energy densities ($H_{BCP}= G_{BCP}+V_{BCP}$), interaction energy ($\Delta E_{HB}= ½\ V_{BCP}$), calculated at the respective bond critical points, by M06-2X2/6-311++G(d,p).

| Compund | $*\rho_{BCP}$ | $*\nabla^2\rho_{BCP}$ | $*V_{BCP}$ | $*G_{BCP}$ | $*H_{BCP}$ | $\Delta E_{HB}$ (kcal/mol) |
|---|---|---|---|---|---|---|
| **$1_{RR}$(OH----O)** | 0.026 | 0.110 | -0.022 | 0.025 | 0.003 | -6.986 |
| **$1_{RR}$(CH----O)** | 0.010 | 0.032 | -0.006 | 0.007 | 0.001 | -1.919 |
| **$1_{RR}^{\cdot+}$(OH----O)** | 0.110 | 0.100 | -0.139 | 0.082 | -0.057 | -43.488 |
| **$1_{RR}^{\cdot+}$(CH----O)** | 0.009 | 0.030 | -0.006 | 0.007 | 0.001 | -1.843 |
| **$1_{SS}$(OH---O)** | 0.026 | 0.110 | -0.022 | 0.025 | 0.003 | -6.990 |
| **$1_{SS}$(CH---O)** | 0.010 | 0.032 | -0.006 | 0.007 | 0.001 | -1.919 |
| **$1_{SS}^{\cdot+}$(OH---O)** | 0.110 | 0.100 | -0.139 | 0.082 | -0.057 | -43.494 |
| **$1_{SS}^{\cdot+}$(CH---O)** | 0.009 | 0.031 | -0.006 | 0.007 | 0.001 | -1.843 |
| **$2_{RR}$(OH----O)** | 0.024 | 0.097 | -0.005 | 0.022 | 0.026 | 1.484 |
| **$2_{RR}$(CH----O)** | 0.007 | 0.020 | -0.004 | 0.004 | 0.001 | -1.256 |
| **$2_{RR}^{\cdot+}$(OH----O)** | 0.084 | 0.146 | -0.098 | 0.067 | -0.031 | -30.62 |
| **$2_{RR}^{\cdot+}$(CH----O)** | ------ | ------ | ------- | ------- | ----- | ------ |
| **$2_{SS}$(OH---O)** | 0.023 | 0.094 | -0.019 | 0.021 | 0.002 | -5.906 |
| **$2_{SS}$(CH---O)** | 0.008 | 0.024 | -0.005 | 0.006 | 0.001 | -1.595 |
| **$2_{SS}^{\cdot+}$(OH---O)** | 0.084 | 0.146 | -0.098 | 0.067 | -0.031 | -30.62 |
| **$2_{SS}^{\cdot+}$(CH---O)** | ------ | ------ | ------- | ------- | ----- | ------ |

*****In a.u.

### 3. A probable justification

Initially, we entertained the following justification for our strange observations. As pure **1*S*** (or **1*R***) passes through the ionization chamber of an EI-MS, each of its six prochiral centers (an -OH and five –$CH_2$ groups) may lose an electron and convert into apparent pairs of diastereomers $\mathbf{1}_{SS}{}^{\cdot+}$ and $\mathbf{1}_{SR}{}^{\cdot+}$ ( or $\mathbf{1}_{RS}{}^{\cdot+}$ and $\mathbf{1}_{RR}{}^{\cdot+}$) with anticipated different energies (Fig. 2a, Supplementary 4). Although, their racemic combination (**1*S*** and **1*R***) don't convert into diastereomers because $\mathbf{1}_{SS}{}^{\cdot+}$ with $\mathbf{1}_{RR}{}^{\cdot+}$ and $\mathbf{1}_{SR}{}^{\cdot+}$ with $\mathbf{1}_{RR}{}^{\cdot+}$ (for species **1*S*** and **1*R***, respectively) are enantiomers. Secondly, discrimination between the mirror images may be justified through their possible self-assemblies as chiral selectors, which could be made diastereomeric complexes *via* diverse interactions such as H-bonding and van der Waals interactions.[58]

The main point seems to be that chiral discrimination might be caused not only by the formation of diastereomeric dimers *via* hydrogen bonding but also by different rates of fragmentation from different conformers of the monomers.[59]

## Conclusion

An unprecedented phenomenon of the ability of chirally blind EI-MS instruments to discriminate between chiral molecules containing prochiral centers is introduced. Our statistical analyses clearly confirm and to an acceptable confidence level significant differences between the EI-MS spectra enantiomers derived from different prochiral compounds (**1*S*** *vs.* **1*R*** and **2*S*** *vs.* **2*R***). The unionized enantiomers (***R*** and ***S***), may act as chiral selectors, making the observed difference between the enantiomers' mass spectra. Hence, self-complexation through H-bonding for **1*S*** (or **1*R***) may give a new chiral center at the H-bonded oxygen atom, producing a new dimeric pair of diastereomers $\mathbf{1}_{SRS}{}^{\cdot+}$ and $\mathbf{1}_{SSS}{}^{\cdot+}$ (or $\mathbf{1}_{RRR}{}^{\cdot+}$ and $\mathbf{1}_{RSR}{}^{\cdot+}$) after ionization. Interestingly, the latter four ionized diastereomers appear with different structural and thermodynamic parameters at the M06-2X/6-311++g (d,p) level of theory. Our statistical analyses clearly confirm and to an acceptable confidence level significant differences between the EI-MS spectra enantiomers derived from different prochiral compounds (**1**$_{\boldsymbol{S}}$ *vs.* **1**$_{\boldsymbol{R}}$ and **2**$_{\boldsymbol{S}}$ *vs.* **2**$_{\boldsymbol{R}}$).

## Conflicts of interest

There are no conflicts to declare.

## Acknowledgments

We thank J.R. Coleman for her technical support, K. Dong for his initial calculations, and Y. Zhang for his initial statistical analyses.

## References:

(1) Andrés-Costa, M. J.; Proctor, K.; Sabatini, M. T.; Gee, A. P.; Lewis, S. E.; Pico, Y.; Kasprzyk-Hordern, B. Enantioselective transformation of fluoxetine in water and its ecotoxicological relevance. *Sci. Rep.* **2017**, *7*(1), 1-13.

(2) Thanzeel, F. Y.; Balaraman, K.; Wolf, C. Click chemistry enables quantitative chiroptical sensing of chiral compounds in protic media and complex mixtures. *Nat. Commun.* **2018**, 9(1), 1-8.

(3) Xu, P.; Liu, Y.; Wang, L.; Wu, Y.; Zhou, X.; Xiao, J.; Zheng, J.; Xue, M. Phencynonate S-isomer as a eutomer is a novel central anticholinergic drug for anti-motion sickness. *Sci. Rep.* **2019**, 9(1), 1-12.

(4) Proctor, R. S.; Colgan, A. C.; Phipps, R. J. Exploiting attractive non-covalent interactions for the enantioselective catalysis of reactions involving radical intermediates. *Nat. Chem.* **2020**, 12(11), 990-1004.

(5) Shen, Q.; Wang, L.; Zhou, H.; Jiang, H. D.; Yu, L. S.; Zeng, S. Stereoselective binding of chiral drugs to plasma proteins. *Acta Pharmacol. Sin.* **2013**, 34(8), 998-1006.

(6) H Brooks, W.; C Guida, W.; G Daniel, K. The significance of chirality in drug design and development. *Curr. Top. Med. Chem.* **2011**, 11(7), 760-770.

(7) Wang, Y.; Yu, Z.; Han, X.; Su, H.; Ji, W.; Cong, Q.; Zhao, B.; Ozaki, Y. Charge-transfer-induced enantiomer selective discrimination of chiral alcohols by SERS. *J. Phys. Chem. C* **2016**, 120(51), 29374-29381.

(8) Ishihara, S.; Labuta, J.; Futera, Z.; Mori, S.; Sato, H.; Ariga, K.; Hill, J. P. NMR spectroscopic determination of enantiomeric excess using small prochiral molecules. *J. Phys. Chem. B* **2018**, 122(19), 5114-5120.

(9) Kalachyova, Y.; Guselnikova, O.; Elashnikov, R.; Panov, I.; Žádný, J.; Církva, V.; Storch, J.; Sykora, J.; Zaruba, K.; Švorčík, V.; Lyutakov, O. Helicene-SPP-based chiral plasmonic hybrid structure: toward direct enantiomers SERS discrimination. *ACS Appl. Mater. Interfaces.* **2018**, 11(1), 1555-1562.

(10) Kumari, D.; Bandyopadhyay, P.; Suryaprakash, N. Discrimination of α-amino acids using green tea flavonoid (-)-epigallocatechin gallate as a chiral solvating agent. *J. Org. Chem.* **2013**, 78(6), 2373-2378.

(11) Young, B. L.; Wu, L.; Cooks, R. G. Mass spectral methods of chiral analysis. Chiral analysis. **2006**, 595-659.

(12) Fales, H. M.; Wright, G. J. Detection of chirality with the chemical ionization mass spectrometer. "Meso" ions in the gas phase. *J. Am. Chem. Soc.* **1977**, 99(7), 2339-2340.

(13) Ahn, S.; Ramirez, J.; Grigorean, G.; Lebrilla, C. B. Chiral recognition in gas-phase cyclodextrin: amino acid complexes—is the three point interaction still valid in the gas phase?. *J. Am. Soc. Mass Spectrom.* **2001**, 12(3), 278-287.

(14) Busacca, C. A.; Senanayake, C. H. 1.9 The use of New Phosphines as Powerful Tools in Asymmetric Synthesis of Biologically Active Compounds. Comprehensive Chirality. Elsevier **2012**, pp. 167-216

(15) Hiep, N. T.; Kwon, J.; Hong, S.; Kim, N.; Guo, Y.; Hwang, B. Y.; Mar, W.; Lee, D. Enantiomeric Isoflavones with neuroprotective activities from the Fruits of Maclura tricuspidata. *Sci. Rep*. **2019**, 9(1), 1-9.

(16) Yu, X.; Yao, Z. P. Chiral recognition and determination of enantiomeric excess by mass spectrometry: A review. *Anal. Chim. Acta.* **2017**, 968, 1-20.

(17) Yang, S.; Wu, F.; Yu, F.; Gu, L.; Wang, H.; Liu, Y.; Chu, Y.; Wang, F.; Fang, X.; Ding, C. F. Distinction of chiral penicillamine using metal-ion coupled cyclodextrin complex as chiral selector by trapped ion mobility-mass spectrometry and a structure investigation of the complexes. *Anal. Chim. Acta.* **2021**, 1184, 339017.

(18) McKendry, R.; Theoclitou, M. E.; Rayment, T.; Abell, C. Chiral discrimination by chemical force microscopy. *Nature* **1998**, 391(6667), 566-568.

(19) Patil, R. A.; Weatherly, C. A.; Armstrong, D. W. Chiral gas chromatography. Chiral Analysis. Elsevier **2018**, pp. 468-505.

(20) Wahab, M. F.; Weatherly, C. A.; Patil, R. A.; Armstrong, D. W. Chiral Liquid Chromatography. Chiral Analysis. Elsevier **2018**, pp. 507-564.

(21) Chankvetadze, B. Enantioseparations by Capillary Electromigration Techniques. Chiral Analysis. Elsevier **2018** pp. 565-605.

(22) Lu, Y.; Wang, Y.; Zhu, W. Nonbonding interactions of organic halogens in biological systems: implications for drug discovery and biomolecular design. *Phys. Chem. Chem. Phys.* **2010**, 12(18), 4543-4551.

(23) Zheng, Y. Z.; Zhou, Y.; Liang, Q.; Chen, D. F.; Guo, R.; Lai, R. C. Hydrogen-bonding interactions between Apigenin and ethanol/water: a theoretical study. *Sci. Rep.* **2016**, 6(1), 1-13.

(24) Loh, C. C. Exploiting non-covalent interactions in selective carbohydrate synthesis. *Nat. Rev. Chem.* **2021**, 5(11), 792-815.

(25) Crawshaw, R.; Crossley, A. E.; Johannissen, L.; Burke, A. J.; Hay, S.; Levy, C.; Baker, D.; Lovelock, S. L.; Green, A. P. Engineering an efficient and enantioselective enzyme for the Morita–Baylis–Hillman reaction. *Nat. Chem.* **2022**, 14(3), 313-320.

(26) Venugopal, P. P.; Das, B. K.; Soorya, E.; Chakraborty, D. Effect of hydrophobic and hydrogen bonding interactions on the potency of ß-alanine analogs of G-protein coupled glucagon receptor inhibitors. *Proteins: Struct. Funct. Genet.* **2020**, 88(2), 327-344.

(27) Singh, S. N.; Yadav, S.; Shire, S. J.; Kalonia, D. S. Dipole-dipole interaction in antibody solutions: correlation with viscosity behavior at high concentration. *Pharm. Res.* **2014**, 31(9), 2549-2558.

(28) Shao, Y.; Molnar, L. F.; Jung, Y.; Kussmann, J.; Ochsenfeld, C.; Brown, S. T.; Head-Gordon, M. Advances in methods and algorithms in a modern quantum chemistry program package. *Phys. Chem. Chem. Phys.* **2006**, 8(27), 3172-3191.

(29) Halgren, T. A. Merck molecular force field. I. Basis, form, scope, parameterization, and performance of MMFF94. *J. Comput. Chem.* **1996**, 17(5-6), 490-519.

(30) Halgren, T. A. Merck molecular force field. V. Extension of MMFF94 using experimental data, additional computational data, and empirical rules. *J. Comput. Chem*. **1996**, 17(5-6), 616-641.

(31) Halgren, T. A.; Nachbar, R. B. Merck molecular force field. IV. Conformational energies and geometries for MMFF94. *J. Comput. Chem*. **1996**, 17(5-6), 587-615.

(32) Gaussian 09, Revision A.02, M. J. Frisch, G. W.; Trucks, H. B.; Schlegel, G. E.; Scuseria, M. A.; Robb, J. R.; Cheeseman, G.; Scalmani, V.; Barone, G. A.; Petersson, H.; Nakatsuji, X.; Li, M.; Caricato, A.; Marenich, J.; Bloino, B. G.; Janesko, R.; Gomperts, B.; Mennucci, H. P.; Hratchian, J. V.; Ortiz, A. F.; Izmaylov, J. L.; Sonnenberg, D.; Williams-Young, F.; Ding, F.; Lipparini, F.; Egidi, J.; Goings, B.; Peng, A.; Petrone, T.; Henderson, D.; Ranasinghe, V. G.; Zakrzewski, J.; Gao, N.; Rega, G.; Zheng, W.; Liang, M.; Hada, M.; Ehara, K.; Toyota, R.; Fukuda, J.; Hasegawa, M.; Ishida, T.; Nakajima, Y.; Honda, O.; Kitao, H.; Nakai, T.; Vreven, K.; Throssell, J. A.; Montgomery, Jr., J. E.; Peralta, F.; Ogliaro, M.;

Bearpark, J. J.; Heyd, E.; Brothers, K. N.; Kudin, V. N.; Staroverov, T.; Keith, R.; Kobayashi, J.; Normand, K.; Raghavachari, A.; Rendell, J. C.; Burant, S. S.; Iyengar, J.; Tomasi, M.; Cossi, J. M.; Millam, M.; Klene, C.; Adamo, R.; Cammi, J. W.; Ochterski, R. L.; Martin, K.; Morokuma, O.; Farkas, J. B.; Foresman, J. Gaussian, Inc., Wallingford CT, **2016**.

(33) Zhao, Y.; Truhlar, D. G. The M06 suite of density functionals for main group thermochemistry, thermochemical kinetics, noncovalent interactions, excited states, and transition elements: two new functionals and systematic testing of four M06-class functionals and 12 other functionals. *Theor. Chem. Acc.* **2008**, 120(1), 215-241.

(34) Zhao, Y.; Truhlar, D. G. Density functionals with broad applicability in chemistry. *Acc. Chem. Res.* **2008**, 41(2), 157-167.

(35) Boys, S. F.; Bernardi, F. J. M. P. The calculation of small molecular interactions by the differences of separate total energies. Some procedures with reduced errors. *Mol. Phys.* **1970**, 19(4), 553-566.

(36) Simon, S.; Duran, M.; Dannenberg, J. J. How does basis set superposition error change the potential surfaces for hydrogen-bonded dimers?. *J. Chem. Phys.* **1996**, 105(24), 11024-11031.

(37) Fang, H.; Kim, Y. Hydrogen-bonded channel-dependent mechanism of long-range proton transfer in the excited-state tautomerization of 7-hydroxyquinoline: a theoretical study. *Theoretical Chemistry Accounts*, **2017**, *136*, 1-13.

(38) Fang, H.; Mai, B.K.; Kim, Y. Excited-State Multiple Proton Transfer Depending on the Acidity and Basicity of Mediating Alcohols in 7-Azaindole–$(ROH)_2$ (R= H, CH 3) Complexes: A Theoretical Study. *Photochemistry and Photobiology*, **2015**, *91*(2), 306-314.

(39) Ikeda, T.; Sakota, K.; Sekiya, H. Elevation of the Energy Threshold for Isomerization of 5-Hydroxyindole-(tert-butyl alcohol) 1 Cluster Cations. *The Journal of Physical Chemistry A*, **2017**, *121*(31), 5809-5816.

(40) Bakker, D.J.; Dey, A.; Tabor, D.P.; Ong, Q.; Mahé, J.; Gaigeot, M.P.; Sibert, E.L.; Rijs, A.M. Fingerprints of inter-and intramolecular hydrogen bonding in saligenin–water clusters revealed by mid-and far-infrared spectroscopy. *Physical Chemistry Chemical Physics*, **2017**, *19*(31), 20343-20356.

(41) Peeters, J.; Nguyen, V.S.; Müller, J.F. Atmospheric vinyl alcohol to acetaldehyde tautomerization revisited. *The Journal of Physical Chemistry Letters*, **2015**, *6*(20), 4005-4011.

(42) Pasechnik, M.P.; Matveeva, A.G.; Lyssenko, K.A.; Aysin, R.R.; Smol'yakov, A.F.; Zubavichus, Y.V.; Godovikov, I.A.; Goryunov, E.I. Competing intramolecular vs. intermolecular hydrogen bonding in phosphoryl-containing secondary alkanols: A structural, spectroscopic and DFT study. *Journal of Molecular Structure*, **2019**, *1175*, 874-881.

(43) Shinkai, T.; Hsu, P.J.; Fujii, A.; Kuo, J.L. Infrared spectroscopy and theoretical structure analyses of protonated fluoroalcohol clusters: the impact of fluorination on the hydrogen bond networks. *Physical Chemistry Chemical Physics*, **2022**, *24*(20), 12631-12644.

(44) Costa, R.J.; Castro, E.A.; Politi, J.R.; Gargano, R.; Martins, J.B. Methanol, ethanol, propanol, and butanol adsorption on H-ZSM-5 zeolite: an ONIOM study. *Journal of Molecular Modeling*, **2019**, *25*, 1-12.

(45) M. H. Kassaee, D. J. Keffer, W. V. Steele, Theoretical Calculation of Thermodynamic Properties of Naphthalene, Methylnaphthalenes, and Dimethylnaphthalenes, *J. Chem. Eng. Data* **2007**, *52*, 1843-1850.

(46) H. T. Wu, D. L. Riggs, Y. A. Lyon, R. R. Julian, Statistical Framework for Identifying Differences in Similar Mass Spectra: Expanding Possibilities for Isomer Identification. *Anal. Chem.* **2023**, *95(17)*, 6996-7005.

(47) Park, S. Y.; Lee, Y. M.; Kwac, K.; Jung, Y.; Kwon, O. H. Alcohol dimer is requisite to form an alkyl oxonium ion in the proton transfer of a strong (photo) acid to alcohol. *Chem. Eu. J.* **2016**, *22(13)*, 4340- 4344.

(48) Fridgen, T. D.; MacAleese, L.; Maitre, P.; McMahon, T. B.; Boissel, P.; Lemaire, J. Infrared spectra of homogeneous and heterogeneous proton-bound dimers in the gas phase. *Phys. Chem. Chem. Phys.* **2005**, *7(14)*, 2747-2755.

(49) Piekara, A. Dielectric saturation and hydrogen bonding. *J.Chem. Phys.* **1962**, *36(8),* 2145-2150.

(50) Eisenschitz, R.; London, F. Über das Verhältnis der van der Waalsschen Kräfte zu den homöopolaren Bindungskräften. *Z. Phys.* **1930**, *60(7)*, 491-527.

(51) Proctor, R. S.; Colgan, A. C.; Phipps, R. J. Exploiting attractive non-covalent interactions for the enantioselective catalysis of reactions involving radical intermediates. *Nat. Chem.* **2020**, *12(11)*, 990-1004.

(52) Su, Z.; Borho, N.; Xu, Y. Chiral self-recognition: Direct spectroscopic detection of the homochiral and heterochiral dimers of propylene oxide in the gas phase. *J. Am. Chem. Soc.* **2006**, *128(51)*, 17126-17131.

(53) Zangi, R. Self-assembly of alcohols adsorbed on graphene. *J. Phys. Chem. C* **2019**, *123(27)*, 16902- 16910.

(54) Grabowski, S. J. (Ed.). Hydrogen bonding: new insights (Vol. 3). Dordrecht: Springer **2006**.

(55) Biegler-Konig, F.; Schonbohm, J.; Bayles, D. AIM2000 - A program to analyze and visualize atoms in molecules. *J. Comput. Chem* **2001**, *22*, 545-559.

(56) Lu, T.; Chen, F. Multiwfn: a multifunctional wavefunction analyzer. *J. Comput. Chem.* **2012**, *33(5)*, 580-592.

(57) Trendafilova, N.; Bauer, G.; Mihaylov, T. DFT and AIM studies of intramolecular hydrogen bonds in dicoumarols. *Chemical physics* **2004**, *302(1-3)*, 95-104.

(58) Scriba, G. K. Recognition mechanisms of chiral selectors: An overview. Chiral Separations. Springer. **2019**, 1-33.

(59) K. B. Wiberg, P. H. Vaccaro, J. R. Cheeseman, (2003). Conformational Effects on Optical Rotation. 3-Substituted 1-Butenes. *J. Am. Chem. Soc.* **2003**, *125(7)*, 1888–1896.